\documentclass[notitlepage,twocolumn,prl,tightenlines,nofootinbib,superscriptaddress,showpacs]{revtex4}

\usepackage{amsmath}
\usepackage{amssymb,amsfonts,latexsym}
\usepackage{bm}
\usepackage[mathcal]{euscript}
\usepackage{graphicx}
\usepackage{epsfig}
\usepackage{color}
\usepackage{pifont}

\definecolor{blue}{rgb}{0,0,1}
\definecolor{bleuf}{rgb}{0,0,0.9}
\definecolor{rougef}{rgb}{0.9,0,0}
\definecolor{green}{rgb}{0,0.5,0}
\definecolor{red}{rgb}{1,0,0}
\definecolor{pink}{rgb}{0.9,0.3,0.7}
\definecolor{azur}{rgb}{0,0.5,0.5}
\definecolor{orange}{rgb}{1,0.5,0.2}
\definecolor{brown}{rgb}{0.5,0,0}

\newcommand{\be}{\begin{equation}}
\newcommand{\ee}{\end{equation}}
\newcommand{\ben}{\begin{equation*}}
\newcommand{\een}{\end{equation*}}
\newcommand{\ba}{\begin{eqnarray}}
\newcommand{\ea}{\end{eqnarray}}

\newcommand{\leg}[1]{\textbf{#1}}

\begin{document}
\graphicspath{{figures/}}

\title{Bulk elastic fingering instability in Hele-Shaw cells}

\author{B. Saintyves}
\email{baudouin.saintyves@cea.fr}
\affiliation{CEA-Saclay, IRAMIS, SPEC, F-91191 Gif-sur-Yvette Cedex, France} 
\author{O. Dauchot}
\affiliation{ESPCI-Paris Tech, PSL$^{\star }$, UMR Gulliver, EC2M, 10 rue Vauquelin, 75231 Paris Cedex 05, France}
\author{E. Bouchaud}
\affiliation{CEA-Saclay, IRAMIS, SPEC,  F-91191 Gif-sur-Yvette Cedex, France}
\affiliation{ESPCI-Paris Tech, PSL$^{\star }$, UMR Gulliver, EC2M, 10 rue Vauquelin, 75231 Paris Cedex 05, France}

\date{\today}

\begin{abstract}
We demonstrate experimentally the existence of a purely elastic, non viscous fingering instability which arises when air penetrates into an elastomer confined in a Hele-Shaw cell. Fingers appear sequentially and propagate within the bulk of the material as soon as a critical strain, independent of the elastic modulus, is exceeded. Key elements in the driving force of the instability are the confinement of the gel and its adhesion to the plates of the cell, which result in a considerable expense of elastic energy during the growth of the air bubble.
\end{abstract}


\maketitle 

Bulk fingering instabilities in viscous liquids confined in Hele-Shaw cells, commonly known as the Saffman-Taylor instability, have given rise to considerable experimental~\cite{Saffman58_ProcRoySoc} and theoretical effort~\cite{Bensimon86_RevModPhys}. In the context of the liquid to solid transition (in gels~\cite{Hirata98_PRE}, foams~\cite{Park94_prl}, yield stress fluids~\cite{Lemaire91_prl,Coussot99_JFM,Lindner00_prl} and Maxwell liquids~\cite{Mora10_PRE}), this instability translates into a fingering to fracture transition.

Purely elastic instabilities in a confined geometry have received much less attention.
An elastic fingering instability has been observed during the peeling of a thin layer of elastomer from a rigid substrate~\cite{Ghatak00_prl,Ghatak06_pre} or in probe tak experiments~\cite{Shull98_MacromolChemPhys} where a semi-spherical probe in contact with the soft solid is pulled up at a constant speed. In most cases, the instability settles on the crack line, which is observed to progress at the interface.~\cite{Ghatak00_prl,Nase08_prl,Adda06_ProcRoySoc,Vilmin10_Langmuir}. 
In a probe-tak geometry, Shull et al.~\cite{Shull00_prl} have observed fingers propagating within the bulk of an acrylic tri-block copolymer gel; but there has been no quantitative characterization of the observed patterns yet, and one dramatically lacks experimental data to constrain the possible mechanisms at the origin of this instability.

\begin{figure}[t!] 
\center
\vspace{0.0cm}
\includegraphics[width=0.5\columnwidth,height=0.5\columnwidth]{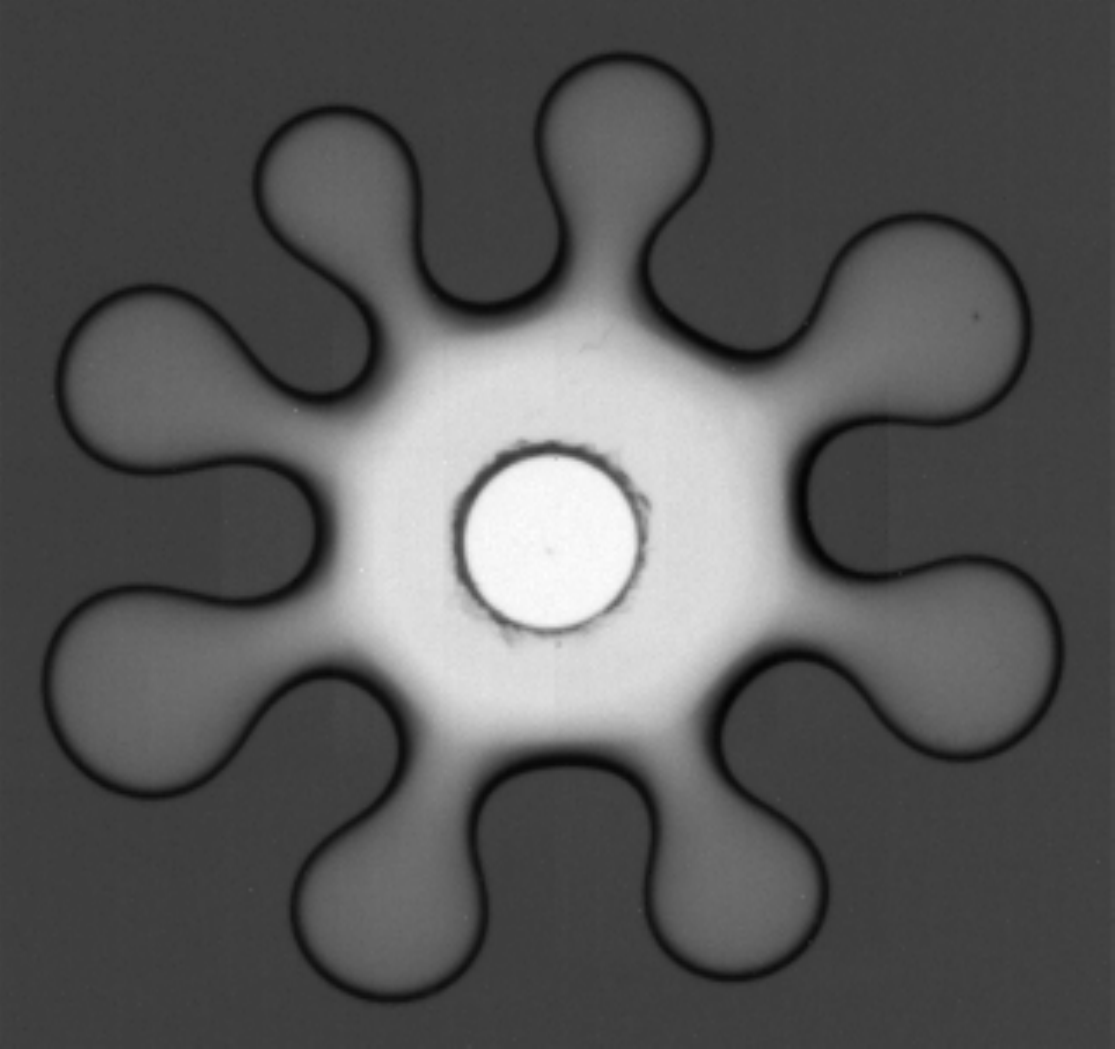}
\vspace{-0.0cm}
\caption{{\bf Elastic fingering instability.} 
Pattern of the destabilized interface ($G=650\,{\rm  Pa}$; $b=2.1 {\rm mm}$; $Q= 126\, {\rm ml/min}$)}
\label{barbabul}
\vspace{-0.4cm}
\end{figure}

In this letter, we demonstrate the existence of a large scale -- centimetric -- elastic fingering instability arising within the bulk of a polyacrylamide gel confined in a Hele-Shaw cell. The instability appears when a critical strain, independent of the shear modulus is exceeded.  Fingers appear sequentially, and, ultimately, lead to the spectacular flower-shaped pattern displayed in Fig.~\ref{barbabul}. The number of fingers increases linearly with the ratio of the initial air bubble diameter to the sample thickness, but it tends to 3 when this ratio tends to zero. More remarkably, the width of the fingers scales in a strongly non linear way with the elastomer thickness.

{\it Materials and methods}.
We use polyacrylamide gels made from acrylamide monomers and bis-acrylamide cross-linkers~\cite{Nossal88_RubChemTech}. The relative concentrations of these two components control the shear modulus of the gel from $G= 80$ to $1060\,{\rm Pa}$. The strain to fracture of the produced materials decreases with $G$, but exceeds 300\% in all cases. Adding dye to the gel allows to see where the material sticks to the glass plates. Cyclic shear rheology of the gels with dye indicates a very small ratio of the elastic modulus to the loss one: $G''/G'<10^{-3}$ for frequencies ranging from $0.01\,{\rm Hz}$ to $100\,{\rm Hz}$.

\begin{figure*}[t!]
\begin{center}
\includegraphics[width=.97 \textwidth] {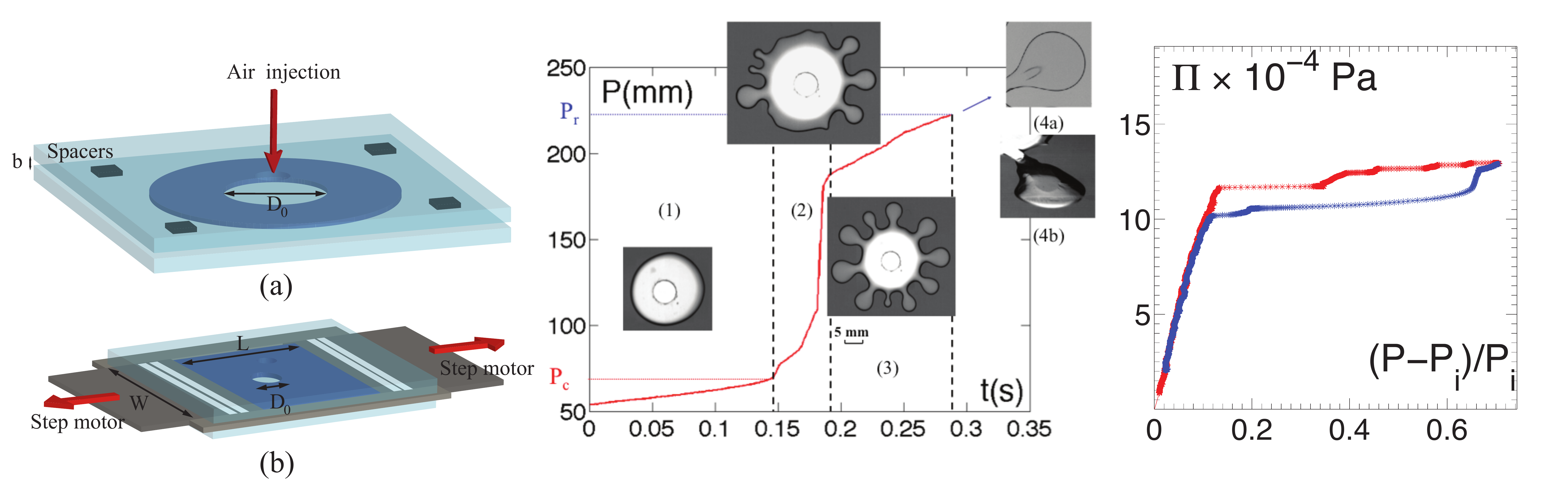}
\caption{{\bf Setups and basic observations:} (color online) \leg{(Left):}  (a): Setup 1, classical Hele-Shaw cell with air injection at the center. (b): Setup 2, sealed Hele-Shaw cell with two moving walls acting as pistons.
 \leg{(Center):}  Sketch of the instability observed in Setup 2 ($G=500\,{\rm Pa}, b=2.1\, {\rm mm}$):  Evolution of the perimeter $P$ of the pattern as a function of time, exhibiting four distinct stages: (1) Circular growth; (2) Nucleation of the fingers; (3) Swelling of the fingers;  (4a) Crack appearing in the gel layer leading to (4b) an interfacial fracture. $P_c$ (respectively $P_r$) is the bubble perimeter at the onset of fingering (reps. just before fracture).  
\leg{(Right):} Reversibility and hysteresis observed in setup 1 ($G=840\,{\rm Pa}, b=1.4\, {\rm mm})$: Pressure $\Pi$ vs. strain $\gamma=(P-P_i)/P_i$, (red curve : inflation; blue curve: deflation).
}
\label{setup}
\end{center}
\vspace{-0.4cm}
\end{figure*}

Two different experimental setups are used. Setup 1 (Fig.\ref{setup}\, left-(a)) is a classical Hele-Shaw cell: it consists in two $10 {\rm mm}$ thick glass plates of lateral sizes $(250 {\rm mm} \times 250 {\rm mm})$ separated by thin spacers of thickness $b\in [0.5-5] {\rm mm}$.  Setup 2 (Fig.~\ref{setup}\, left-(b)) is an original design, which consists in a sealed Hele-Shaw cell made of two $10 {\rm mm}$ thick glass plates of lateral sizes $(250 {\rm mm} \times 125 {\rm mm})$, with two opposite mobile walls, acting as pistons,  and pulled at a prescribed velocity $V$ by synchronized step motors. The gap is fixed to the value $b=2.1{\rm mm}$. This setup is closer to classical tensile tests in solid mechanics.
In both experiments the cells are filled with polyacrylamide before gelation. During this process, we maintain an initial air bubble of controlled diameter $D_0$.  In Setup 1, air is blown with a  syringe pump in order to grow this bubble. The pressure varies from 0 to 1.5 bar. 
In Setup 2, the depression induced by the motion of the pistons sucks air into the gel. Because polyacrylamide is incompressible, the speed and size of the pistons impose an air flow rate equal to $Q=2WbV= 126 {\rm ml/min}$.  In both cases, the 
loading rates are such that the material can safely be considered as purely elastic. The temperature was kept constant, at 21.5C, during all the experiments (including rheology).

Most observations are top views of the cell recorded at $50 {\rm Hz}$ for Setup 1 and $1500 {\rm Hz}$ for Setup 2, which gives an idea of the robustness of the instability over a wide range of interface speeds. We measure the thickness of the gel layers through light absorption. When performing such measurements, we use a silicon oil (Rhodorsil V20), which is immiscible with the gel but has a similar refractive index and therefore avoids refraction. The observed patterns show no differences with the one obtained when injecting air: The interfacial tension between the injected fluid and the gel is not a relevant parameter.

{\it Experimental results.}
The same scenario holds in both setups and can be summarized as follows (Fig.~\ref{setup}-center and Supp. Materials~\cite{Supp}). The evolution of the perimeter $P$ of the pattern allows to distinguish four stages. During stage (1), a circular bubble grows. 
Shades of gray behind the dark line of the front show the existence of layers of gel adhering to the glass plates. In stage (2), fingers burst out successively, with an experiment-dependent order of appearance, and with a speed much larger than the velocity of circular extension. This leads to a sharp increase of the perimeter. Stage (3) corresponds to the swelling of the fingers once they are all formed, and the increase of the perimeter is much slower again. Then, in stage (4), the layer of gel behind the tip breaks, leading to an interfacial crack.

Prior to the interfacial fracture stage, the phenomenon is completely reversible: Fingers deflate when the tensile stress is released, and the initial bubble is recovered. The reversibility of the process clearly demonstrates the purely -- possibly non linear --  elastic nature of the instability. Figure~\ref{setup} (right) displays the pressure $\Pi$ as a function of strain $\gamma=(P-P_i)/P_i$, where $P_i=\pi D_0$ is the initial bubble perimeter. It shows that the phenomenon is hysteretic: the pressure needed to form the fingers is significantly larger than the pressure released when they deflate. When air is injected a second time, the pressure follows the red curve again, which shows that the material has not been damaged during the first cycle.

 \begin{figure}[t!]
\begin{center}
\includegraphics[width=0.98\columnwidth,height=.5 \columnwidth] {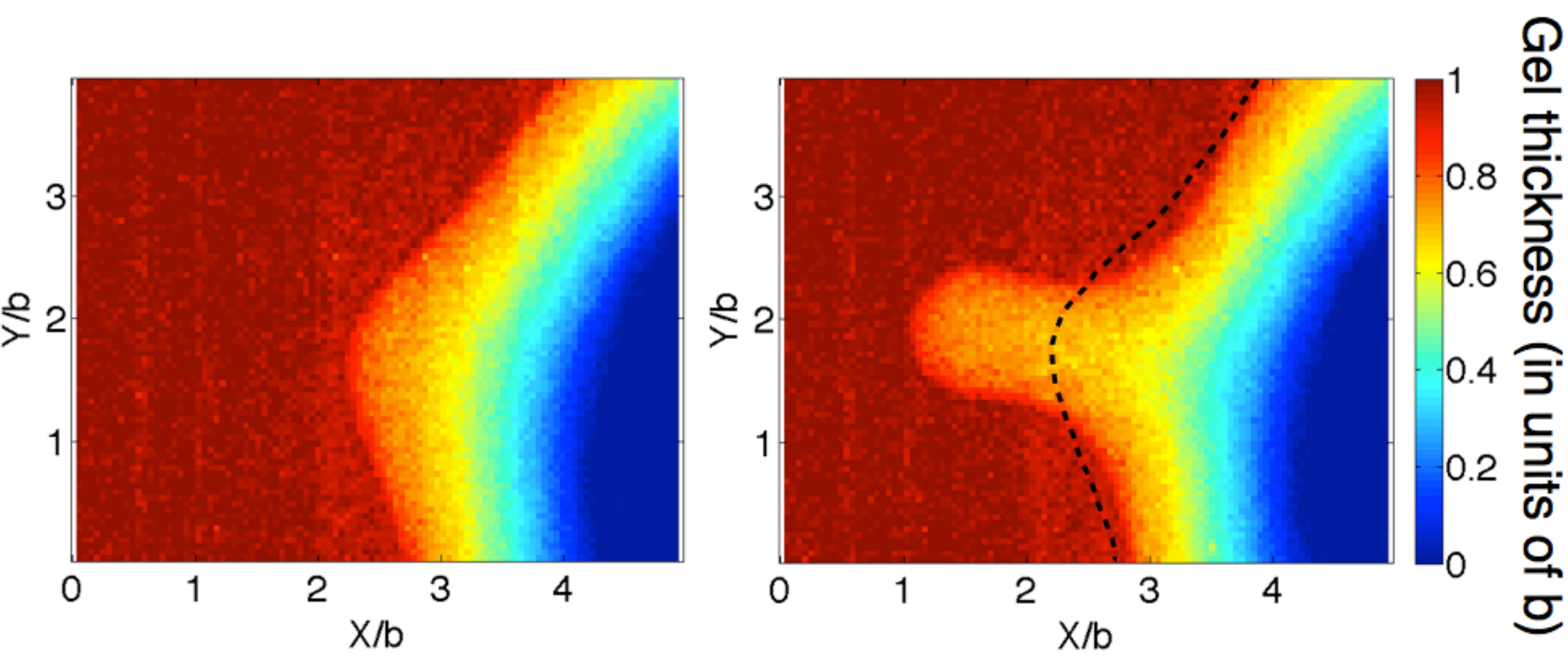}
\vspace{-2mm}
\hspace{0.5\columnwidth}(a)\\
\includegraphics[width=0.95\columnwidth] {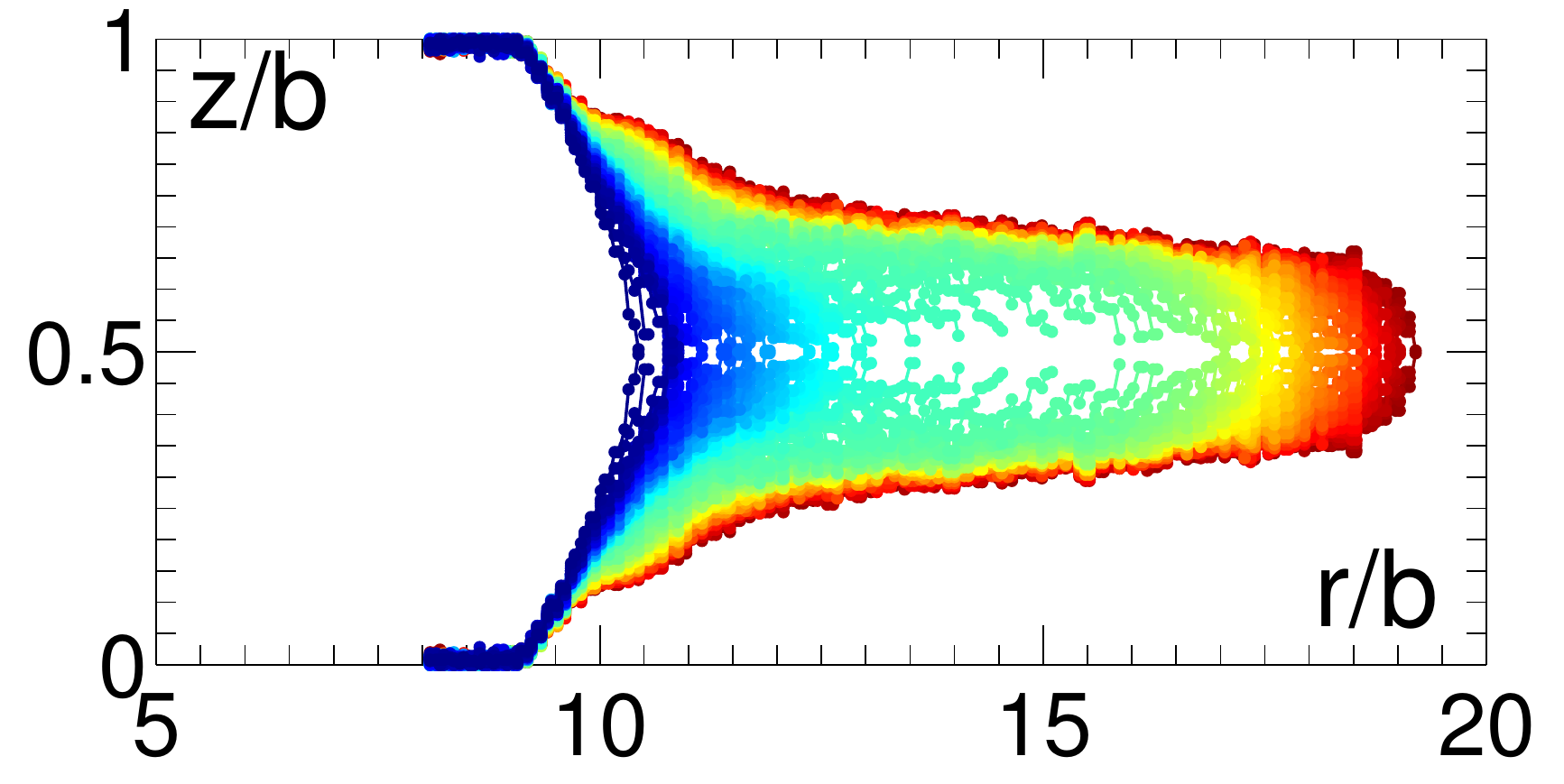}
\vspace{-3mm}
\hspace{0.5\columnwidth}(b)\\
\caption{{\bf Formation of a finger:} (color online) \leg{(a):}Top view before (left) and after (right) the finger has formed. The initial front (black dotted line) recedes in the finger vicinity. Color bar: thickness of the gel in $b$ units. \leg{(b)} : Thickness profiles of a finger developing in time. Color codes 137 time steps from blue to red separated by $0.6\,{\rm ms}$. $G = 100\,{\rm Pa}$.}
\label{thickness}
\end{center}
\vspace{-0.6cm}
\end{figure}

By filming the cell at an angle, we could see a clear meniscus on all fingers. Besides, as already shown in Fig.~\ref{setup}-(center), interfacial cracking occurs at a late stage, after the full development of the instability. There, two distinct fronts are easily observable, one corresponding to the interfacial crack, the other one to a bulk deformation ahead of it (Fig.~\ref{setup}-(center) image 4(b)).  Further quantitative evidence of this 3D character is provided in Fig.~\ref{thickness}, where color encodes the gel thickness measured from light absorption (Fig.~\ref{thickness}). 
Before a finger bursts out, the interface is already locally deformed. The gel thickness increases with the distance to the injection point. The finger first develops with a constant width and almost with a constant thickness as further illustrated in Fig.~\ref{thickness}(b), where the finger profiles within the thickness of the cell have been plotted at constant time intervals, assuming a symmetric shape. The finger grows up to a certain size where only the extremity is swelling while the back, forming a saddle, does not evolve any more. This early stage is followed by a regime where both the extremities and the ``saddle'' parts of the fingers increase in size.  Note that the formation of the finger is accompanied by a relaxation of the elastic strain in its vicinity, where the front is seen to recede (Fig.~\ref{thickness}(a)-right). The two phenomena arise instantaneously - within experimental accuracy -, and are far more rapid than the growth rate of the whole pattern. In some cases, one observes small fingers, which start to develop and which eventually recede, if they are surrounded by larger ones (see inset (3) of Fig.~\ref{setup}-(center). Again this is a clear signature of the elastic nature of the process.

\begin{figure}[t!] 
\center
\vspace{0.0cm}
\includegraphics[width=.49 \columnwidth] {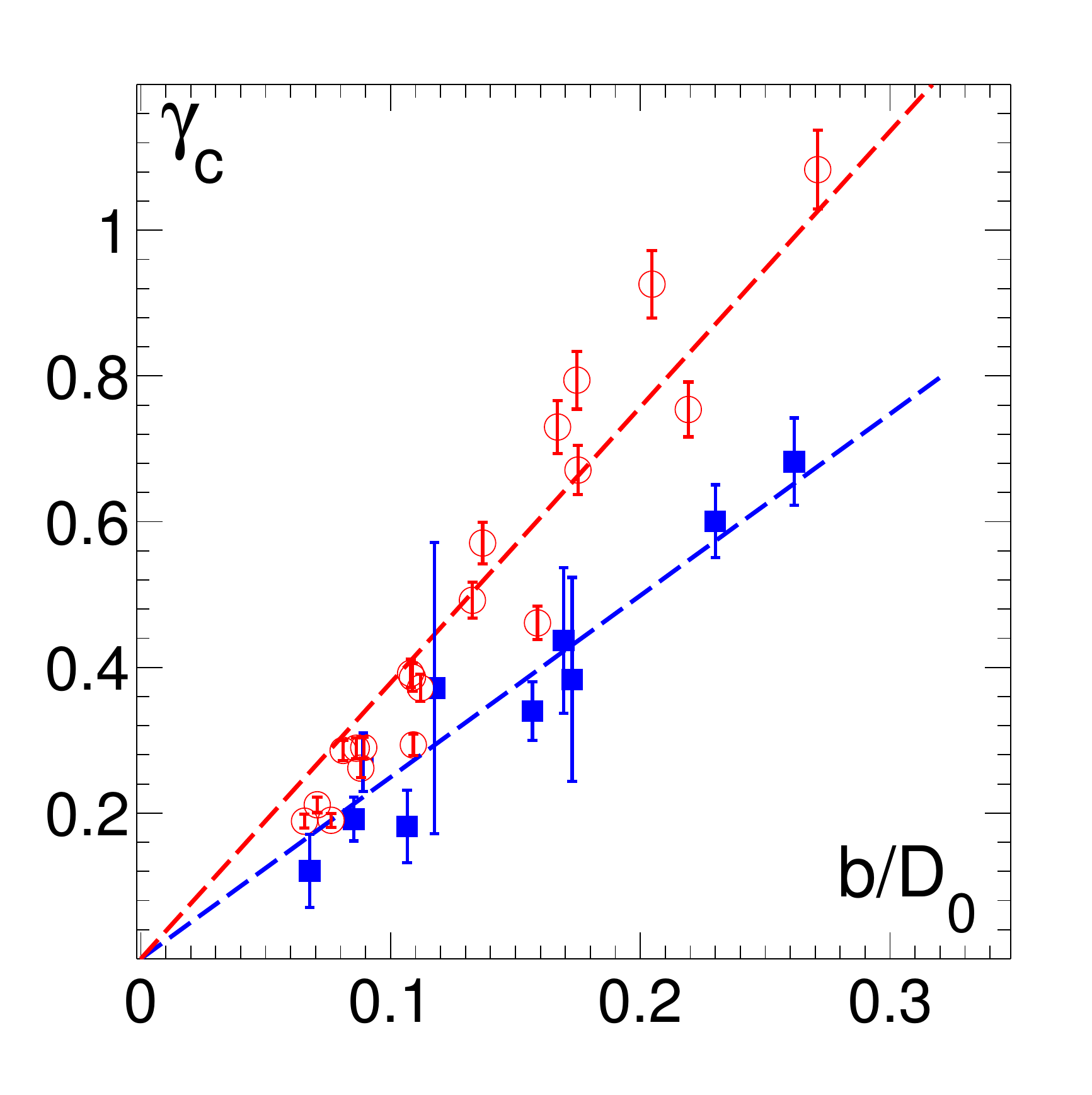}
\includegraphics[width=.49 \columnwidth] {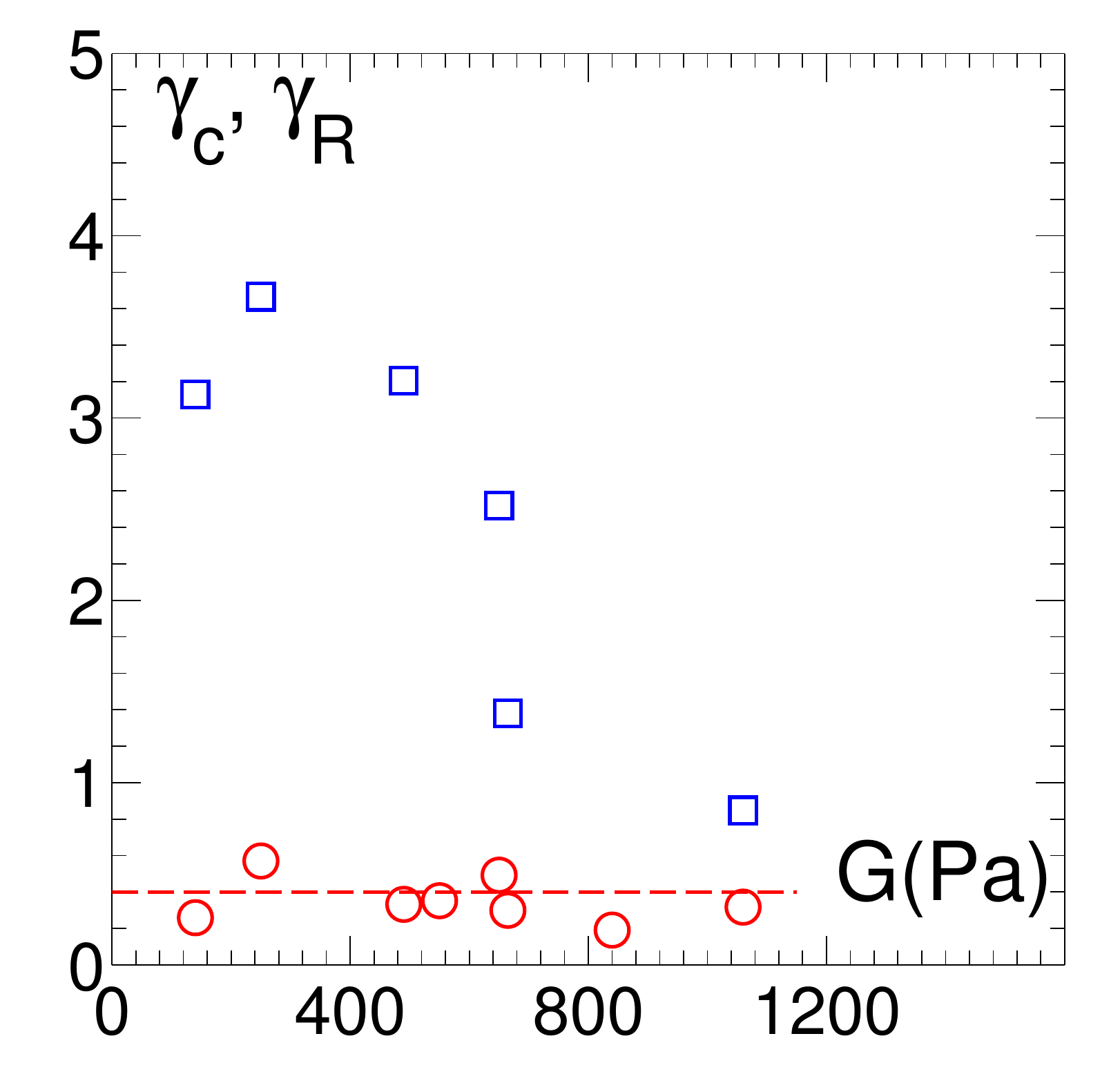}
\vspace{-2mm}
\hspace{0.05\columnwidth}(a)\hspace{0.49\columnwidth}(b)\\
\vspace{-2mm}
\caption{{\bf Critical strain} (color online):
\leg{(a):}  $\gamma_c$ at the onset of fingering as a function of $b/D_0$ for Setup1 (\textcolor{blue}{$\blacksquare $}) and Setup 2 (\textcolor{red}{o}). \leg{(b):}  Setup 2- critical strain $\gamma _c$,$(\textcolor{red}{\circ})$, and strain to fracture $\gamma _R, (\textcolor{blue}{\square})$, as a function of the shear modulus $G$, $D_0=23mm$.}
\label{shearmod}
\vspace{-0.4cm}
\end{figure}

As can be seen from the perimeter temporal evolution (Fig.~\ref{setup}-center), the instability sets in when the strain exceeds a critical value $\gamma_c=(P_c-P_i)/P_i$ which increases linearly with  $b/D_0$ for both setups (see fig.~\ref{shearmod}(a)). The mechanism of the instability is the same for the two experiments, but the stress distribution depends on the geometry of the device, hence the difference of slope. It is remarkable that $\gamma_c$ does not depend on the shear modulus (Fig.~\ref{shearmod}(b)), in contrast with the strain to fracture $\gamma_r=(P_r-P_i)/P_i$, which decreases with $G$. 

 \begin{figure}[t!]
\begin{center}
\includegraphics[width=.49 \columnwidth] {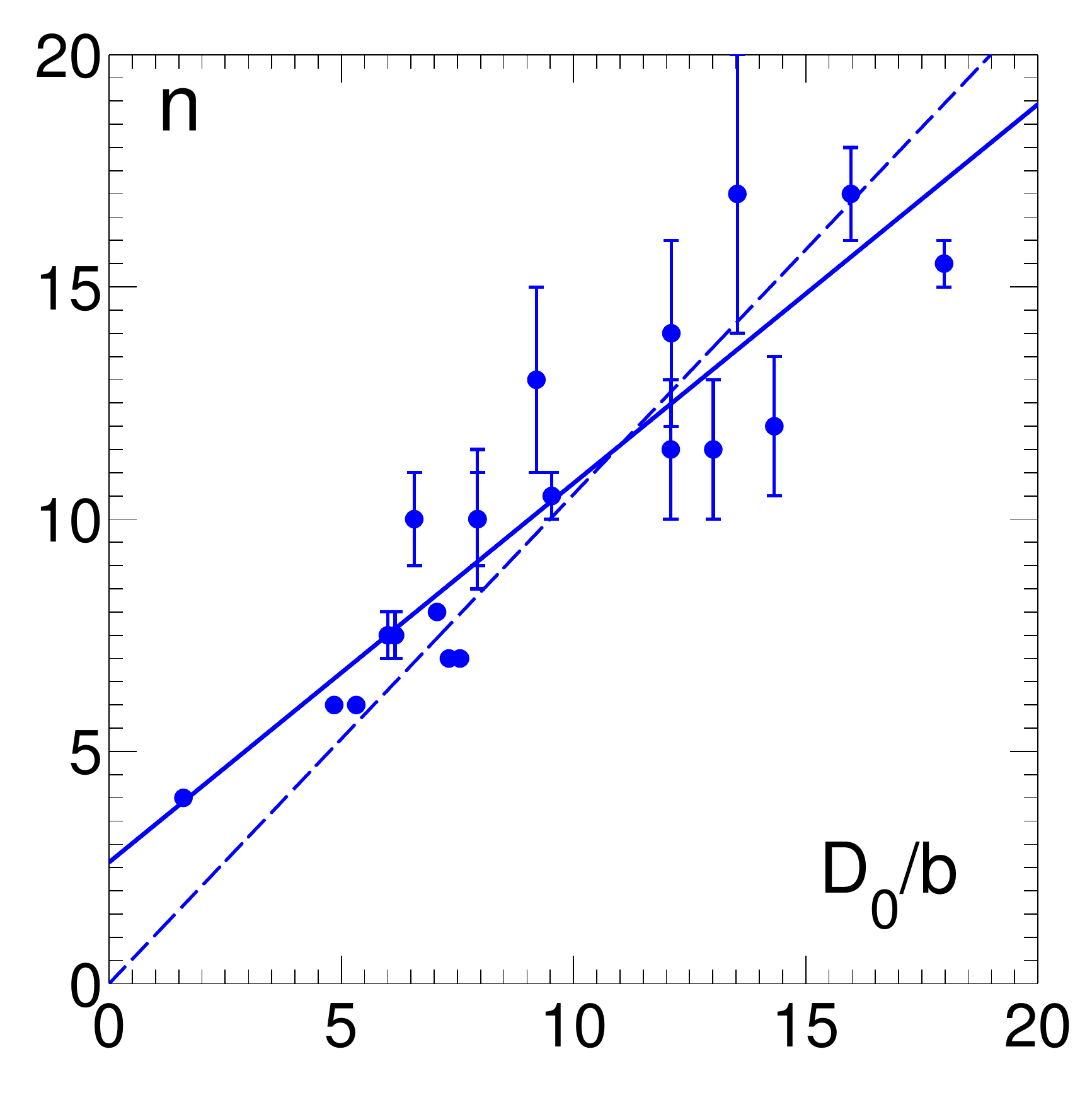}
\includegraphics[width=.49 \columnwidth] {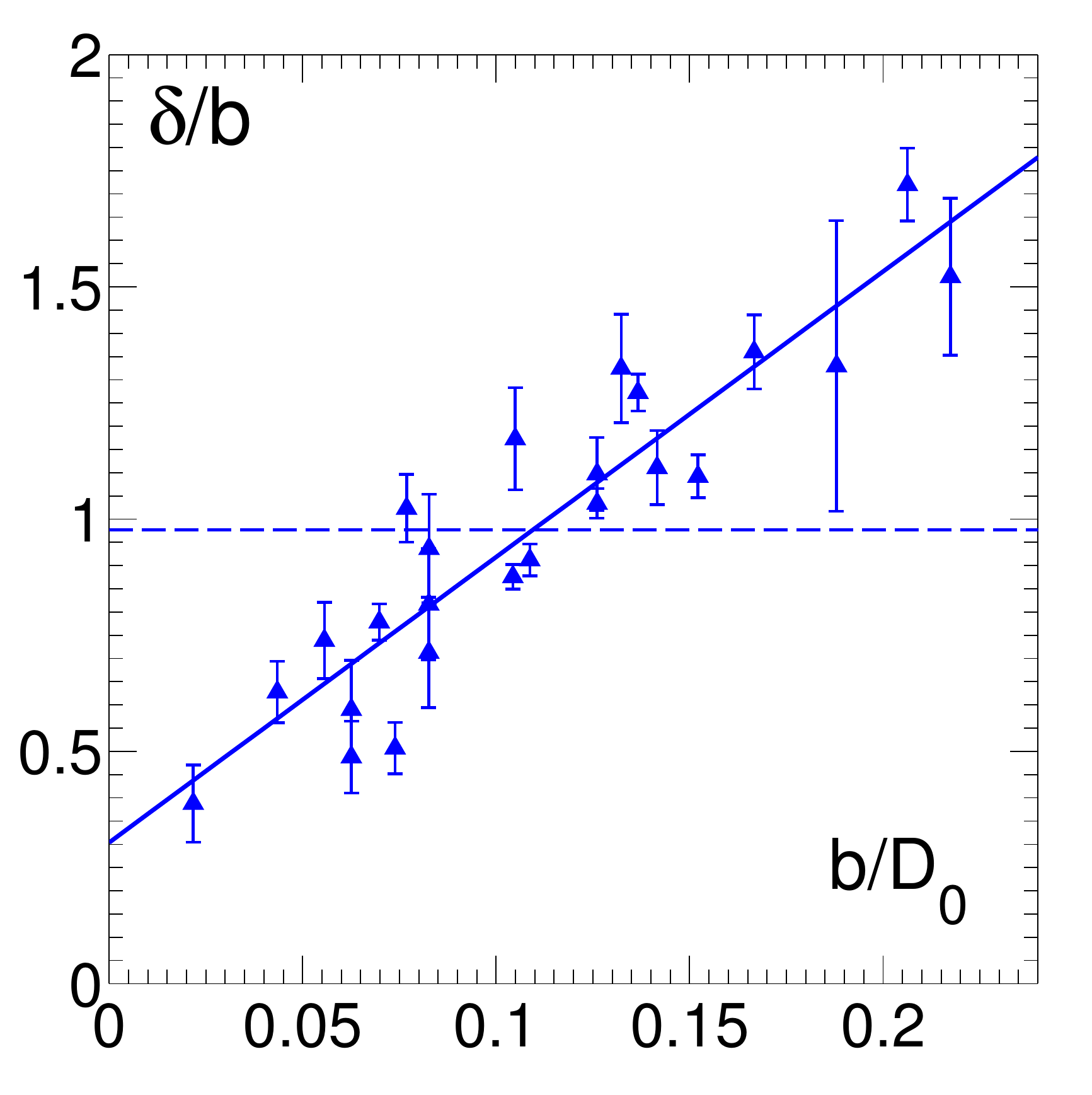}
\vspace{-1mm}
\hspace{0.05\columnwidth}(a)\hspace{0.49\columnwidth}(b)\\
\vspace{-2mm}
\caption{{\bf Influence of the ratio $D_0/b$  (Setup 1)} (color online) 
\leg{(a):} Number of fingers $n$ as a function of $D_0/ b$. The black dashed line is a linear fit; the solid black line is an affine fit extrapolating to $n=3$ when $D_0/ b$ tends to 0; \leg{(b):}  $\delta /b$ - width of the fingers $\delta $ at their roots in $b$ units as a function of $b/D_0$.
The blue dash line is the simplest expected scaling, here a constant. The solid line is an affine fit (see text for details).}
\label{gap}
\end{center}
\vspace{-0.6cm}
\end{figure}

Finally experiments performed with Setup 1, for which $b$ can be varied easily, allow to estimate the role of the vertical confinement. The number of fingers $n$ (Fig.~\ref{gap}(a)) and the width $\delta/b$ of the fingers at their roots in units of $b$, measured when they are just formed (Fig.~\ref{gap}(b)) depend only on $D_0/b$. Because $\delta $ is not a wavelength, there is no simple relationship between $\delta $, $n$ and $D_0$.
The expected scaling $n\propto D_0/b$ (dash line in Fig.~\ref{gap}(a)) is not incompatible with our data, although they are better fitted by an affine law suggesting a finite number of fingers ($n\simeq 3$) when $D_0/b$ tends to zero. 
More strikingly, the fingers width $\delta/b$ is clearly not a constant (dashed line in Fig.~\ref{gap}(b)) as one could have expected; $\delta/b$ rather follows an affine law (solid line in Fig.~\ref{gap}(b)), which tends to a constant ($\simeq 0.3$) when $b/D_0$ tends to zero. This suggests that the simple expected scaling $\delta \propto b$ is verified only when $b$ tends to zero (infinitely thin layer of elastomer) or $D_0$ tends to infinity (limit of an initial straight front).

\begin{figure}[t!]
\begin{center}
\includegraphics[width=.535 \columnwidth,height=.47 \columnwidth] {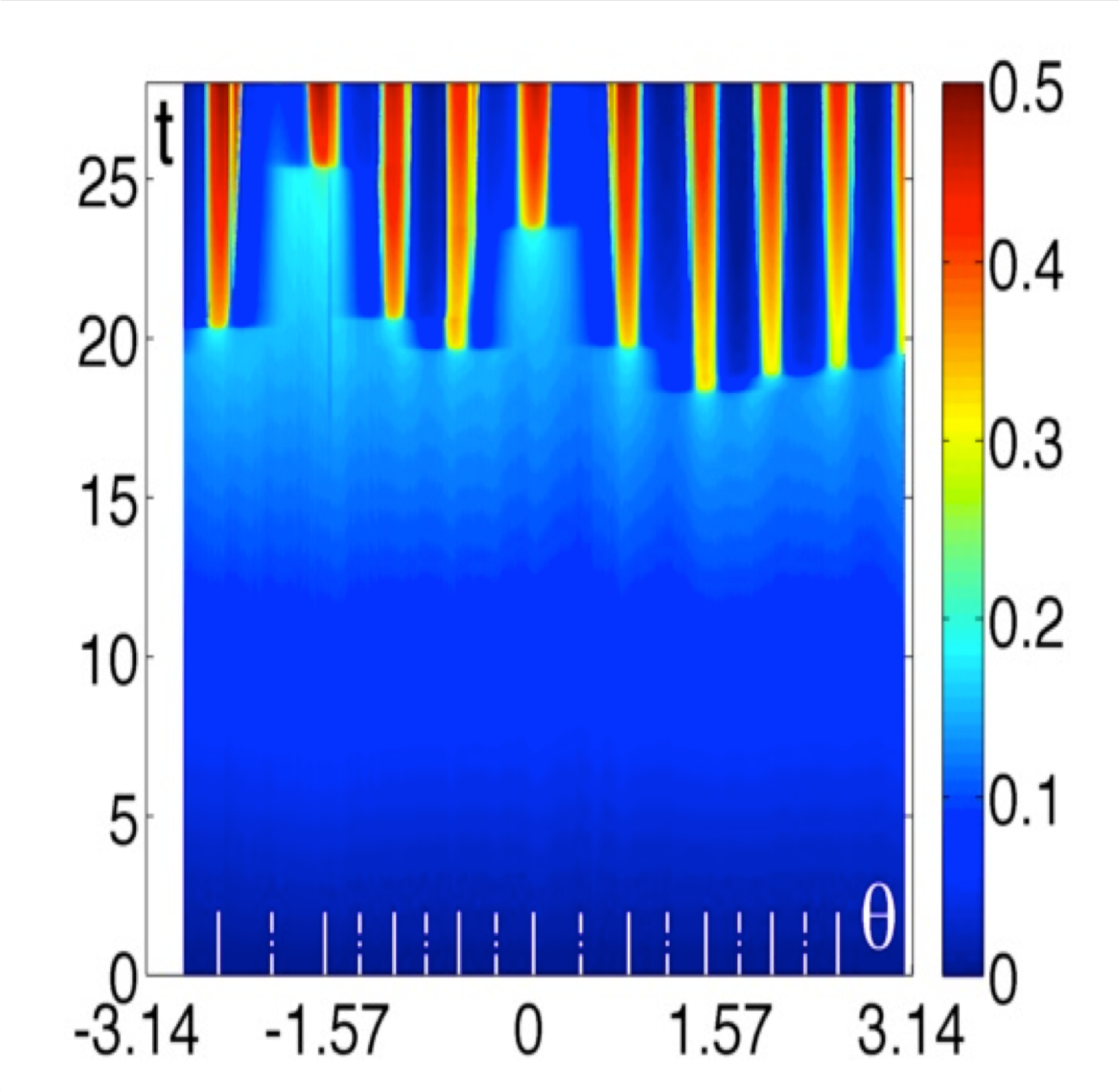}
\includegraphics[width=.45 \columnwidth,height=.47 \columnwidth] {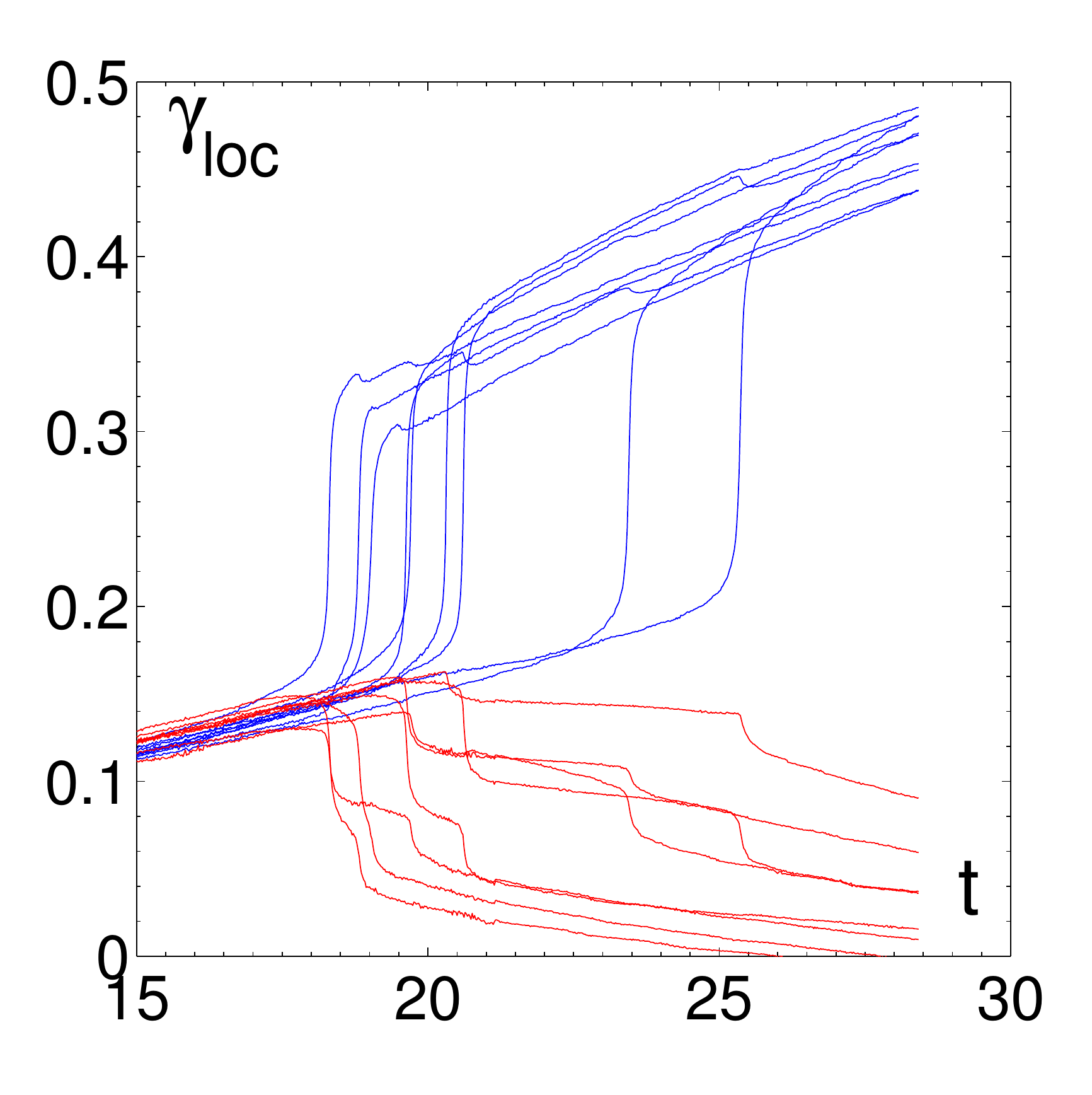}
\vspace{-2mm}
\hspace{0.05\columnwidth}(a)\hspace{0.44\columnwidth}(b)\\
\vspace{-2mm}
\caption{{\bf Spatio-temporal diagram :} (color online) 
\leg{(a):}  Local strain $\gamma_{loc}(\theta,t)$; \leg{(b):}  Evolution of $\gamma_{loc}$ as a function of time: $-$ {\it Blue curves}: along the fingers (indicated by ticks in solid white lines in (a)); $-$ {\it Red curves}: between two adjacent fingers, (as indicated by ticks in dotted-dashed white lines in (a)). $b=2.9\,{\rm mm}$.}
\label{gap2}
\end{center}
\vspace{-0.6cm}
\end{figure}

Fig.~\ref{gap2}(a) displays a spatio-temporal diagram of the local strain $\gamma_{loc}=(r(\theta,t)-r(\theta,0))/r(\theta,0)$ recorded in Setup 1 for a gap of $2.9\, {\rm mm}$. Despite their sequential appearance, fingers end up regularly spaced, without having drifted. One can even identify the location of a missing finger ($\theta \simeq -2\pi /3$ in Fig.~\ref{gap2}(a)), which would eventually develop, if the pattern were not to fracture before. This suggests the existence of a well defined wavenumber. Fig.~\ref{gap2}(b) shows the evolution of strain both along each finger, and between the fingers. The strain is evenly distributed until the fingers start to grow. Then blue curves take off, showing stress concentration at each finger tip, while red curves decrease, showing that this local stress concentration is accompanied by a stress release in between fingers. This release is what causes the front to recede in the close vicinity of fingers.

{\it Discussion.}
We have reported experimental evidence of a bulk fingering instability arising in a confined layer of a purely elastic gel. This bulk instability is clearly different from the interfacial fingering observed in peeling~\cite{Ghatak00_prl, Ghatak06_pre}. It shares some similarities with the preliminary probe tak experiments reported by  Shull et al.~\cite{Shull00_prl}, but a crucial difference with their observations lies in the existence of a clear hysteresis in the present case.

The origin of the instability reported here is, we believe, very similar to the well known balloon instability in hyper-elastic materials~\cite{Gent05_NLM}. In that case, the 
dependence of the pressure inside the balloon as a function of its radial elongation has a non-monotonic shape. While for a pressure lower than a certain value, there is a unique elongation, for larger pressures there are three solutions for a given value of the balloon radius. The solution which corresponds to the intermediate deformation being unstable, the deformation jumps from the smallest to the largest value. 
In a very similar way, the layers of gel which stick to the glass plates are sheared due to the inflation of the air bubble, up to a point where they can elongate considerably at constant stress. Indeed, if one considers that, due to confinement, stresses are screened out at distances of the order of the gap $b$, then one can restrict the analysis to a membrane of gel of initial thickness $b$, of cylindrical shape, with internal radius $R_0$ and height $b$. An increase $\Delta R \gg b$ of the bubble radius results into an elongation of this membrane of order $1+\Delta R/R_0 $ in the orthoradial direction, and $2\Delta R /b$ in the radial direction. As a consequence, because of the gel incompressibility, the final thickness of the membrane  is equal to $b/[2(\Delta R/b)(1+\Delta R/R_0)]$. Using Gent's hyperelastic model ~\cite{Gent96_RubChemTech,Gent05_NLM} to link stresses and elongations, one finds that the pressure within the bubble increases up to a maximum value as a function of $\Delta R/R_0 $ and reaches a plateau before increasing again at much larger deformations.

  As in the balloon instability, the physical origin of the observed phenomenon resides in the fact that, beyond a certain pressure, it becomes easy to stretch the two hyperelastic films that adhere to the surfaces of the cell. In this sense, the observed instability is very generic, and its origin resides (i) in the hyperelasticity of the material, (ii) its incompressibility and (iii) the non-gliding boundary conditions at the glass plates. As a matter of fact, experiments performed on an agar gel which glides along glass surfaces show no instability. On the contrary, the same elastic instability has been observed also in connected microemulsions which stick to glass like polyacrylamide. 

A more detailed mathematical analysis, that partially accounts for the results reported here, can be found in~\cite{Biggins13_PNAS}. However, much 
remains to be done to deepen and exploit the balloon instability analogy.

Finally, let us recall that fingering is commonly associated to liquids, while fracture is associated to solids. In this experiment, we show that fingering occurs also within the bulk of a purely elastic soft confined solid.  In future work, it will be particularly interesting to analyze the crossover from our instability to a viscous fingering instability in a viscoelastic medium, as a function of the injected flow rate. 

{\bf Acknowledgements:} Nothing could have been done without the technical support of V. Padilla and C. Gasquet. This work was funded by ANR (F2F project) and by Triangle de la Physique (FracHet). Many thanks to F. Boulogne for his hospitality at FAST/Orsay, and to R. Gy for providing plates made of Saint-Gobain float glass. We are also indebted to S. Mora for interesting discussions. Particular thanks to J.-P. Bouchaud for enlightening suggestions, and to J. Biggins and L. Mahadevan, who have suggested a theoretical model for this instability. E.B. also acknowledges the hospitality of CAS in Oslo, Norway.

\vspace{-4mm}
\bibliography{Baudouin_ref2.bib}
\end{document}